# PGA Tour Scores as a Gaussian Random Variable

Robert D. Grober
Departments of Applied Physics and Physics
Yale University, New Haven, CT  06520

Abstract

In this paper it is demonstrated that the scoring at each PGA Tour stroke play event can be reasonably modeled as a Gaussian random variable. All 46 stroke play events in the 2007 season are analyzed. The distributions of scores are favorably compared with Gaussian distributions using the Kolmogorov-Smirnov test. This observation suggests performance tracking on the PGA tour should be done in terms of the *z*-score, calculated by subtracting the mean from the raw score and dividing by the standard deviation. This methodology measures performance relative to the field of competitors, independent of the venue, and in terms of a statistic that has quantitative meaning. Several examples of the use of this scoring methodology are provided, including a calculation of the probability that Tiger Woods will break Byron Nelson's record of eleven consecutive PGA Tour victories.

Statistical analysis is now a ubiquitous aspect of most professional sports [1]. Perhaps the best example of this is professional baseball, where nearly every aspect of the game is framed in terms of statistical analysis [2]. Professional golf is also a sport that focuses intensely on statistics, as the PGA Tour web site (www.pgagtour.com) maintains statistics on many aspects of the performance of individual players. The goal of this paper is to demonstrate that the 18-hole scores reported on the PGA tour are reasonably described in terms of Gaussian statistics. The scores generated by the field of competitors at each venue is characterized in terms of a mean, $\mu$, and variance, $\sigma^2$, and the histogram of scores at each venue is accurately described by the associated Gaussian probability distribution function.

The central limit theorem states that if the random variable $y$ is the sum over many random variables $x_i$, $y = \sum_{i=1}^{N} x_i$, where $N \gg 1$, then $y$ will be a Gaussian distributed random variable, *i.e.* the probability density function for $y$ is given as

$$p(y) = \frac{1}{\sqrt{2\pi\sigma_y^2}} \exp\left[-\frac{(y-\mu_y)^2}{2\sigma_y^2}\right]$$

where $\mu_y = \langle y \rangle$, $\sigma_y^2 = \langle y^2 \rangle - \langle y \rangle^2$, and the brackets $\langle \ \rangle$ denote an average. Thus,

$\mu_y = \sum_{i=1}^{N} \langle x_i \rangle = \sum_{i=1}^{N} \mu_{x_i}$. In the limit the random variables $x_i$ are uncorrelated,

$\sigma_y^2 = \sum_{i=1}^{N} \left(\langle x_i^2 \rangle - \langle x_i \rangle^2\right) = \sum_{i=1}^{N} \sigma_{x_i}^2$, and thus $\mu_y$ and $\sigma_y^2$ are defined by the first and second moments of the random variables $x_i$ [3].

The central limit theorem is independent of the statistics of $x_i$ so long as $N \gg 1$. However, in the limit that the $x_i$ are well behaved (i.e. not so different than Gaussian), $N$ need not be very large before $y$ is Gaussian. In the limit that the $x_i$ are Gaussian, the central limit is trivially true even for $N \sim 1$, as the sum over Gaussian random variables is a Gaussian random variable.

The final score of a round of golf is defined by the sum $y = \sum_{i=1}^{N} x_i$, where $x_i$ is the score on the $i^{th}$ hole and $N = 18$. One might reasonably expect the central limit theorem to be relevant to the distribution of scores at a golf tournament provided the distribution of strokes taken on each hole, $x_i$, are reasonably behaved random variables. This is likely the case for tournaments involving golf professionals, who generally score not too much different from par on each hole.

As a preliminary test of this hypothesis, the scores reported for the 2007 PGA Tour Qualifying School at Orange County National, November 28, 2007 through December 3, 2007 were analyzed. The event involved 158 golfers playing six rounds of tournament golf. Each golfer played three rounds of golf on each of the two different courses, Panther Lake Course and Crooked Cat Course. The final results posted online by the PGA tour [4] do not distinguish between the two courses. Likewise, the analysis does not distinguish between the two courses.

The probability distribution of the 948 scores is shown in Fig. 1. This distribution is calculated by first making a histogram of all the scores. The probabilities are calculated from the histogram by normalizing the number of counts in each bin by the total number of scores. The uncertainties are estimated as the square root of the number of scores in a particular bin normalized to the total number of scores, *i.e.* it is assumed the statistics associated with the number of scores in each bin are Poisson [5]. The resulting probability distribution and estimated uncertainties are indicated by the vertical bars in Fig. 1. The first and second moment of the probability distribution of scores yield a mean $\mu_s = 70.8$ strokes and standard deviation $\sigma_s = 2.6$ strokes.

Motivated by the central limit theorem, a model for this probability distribution is shown as the solid line in Fig. 1. This model distribution is obtained numerically by taking $10^5$ samples of a Gaussian random variable with the same mean and standard deviation as the data, then rounding each sample to the nearest integer. Visually, the model seems to be a very reasonable representation of the data.

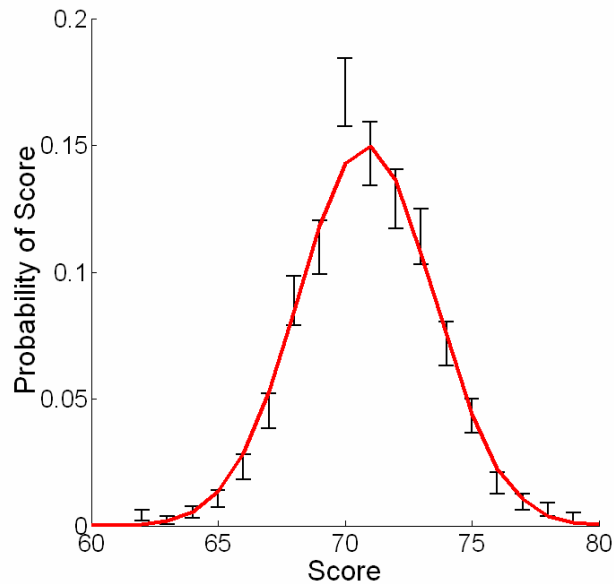

Figure 1: The probability distribution of the 948 scores reported for the 2007 PGA Tour Qualifying School. Calculation of the first and second moment yield a mean $\mu_s = 70.8$ strokes and standard deviation $\sigma_s = 2.6$ strokes. The vertical bars represent the probability distribution and estimated uncertainty. The solid line is the model distribution. All calculations are described in the text.

A standard test of the similarity of two sample populations is the Kolmogorov-Smirnov (*K-S*) test [6], in which the cumulative distribution functions of the two sample populations are compared. The *K-S* test is designed to test as a null hypothesis whether the two distributions are drawn from the same underlying distribution. This type of testing is particularly sensitive to errors in the tails of the probability distribution functions. The resulting *p*-value defines the significance level at which one can reject the null hypothesis. It is standard practice to reject the null hypothesis at significance levels five percent or less.

The *K-S* test is based on an analysis of the Quantile-Quantile (*QQ*) plot, comparing the cumulative distribution functions of the two sample populations. One representation of the *QQ* plot is shown in Fig. 2, in which the inverse of the cumulative distribution function (*CDF*) for both sample populations, the data and the model, is calculated in 1% increments from 1% through 100%, yielding 100 data points for each sample population. These two data sets are then plotted parametrically with the model on the ordinate axis and the data on the abscissa axis. Each data point is represented by a cross in Fig. 2. Because many of the points are overlapping, each data point in the graphic has been dithered by the addition of a random number drawn from a Gaussian distribution with $\mu = 0$ strokes and standard deviation $\sigma = 0.2$ strokes. The solid line shown in Fig. 2 is the line $x = y$, which is what would be observed for two identical sample populations. The *QQ* plot is very linear, indicating the two sample populations are similar.

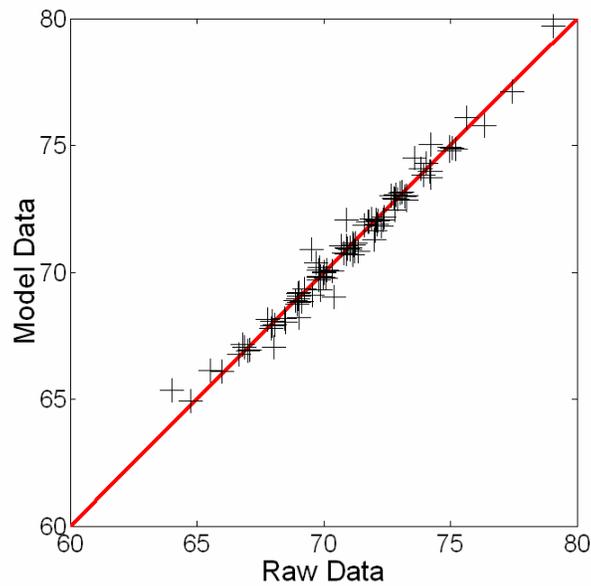

Figure 2: A *QQ* plot of the raw data and the model data, indicated as crosses. For purposes of presentation, each data point has been dithered randomly by a small amount so that the density of data points can be appreciated. The solid line is the line $x = y$, which is what would be observed for two identical distributions. The *K-S* test returns a *p*-value of 0.92, making it difficult to reject the null-hypothesis that the two samples are drawn from the same distribution.

The *K-S* test is performed using a numerical analysis provided in MATLAB [7]. Because the resulting *p*-value is 0.92, the null hypothesis can not be rejected. Thus, it is convincing that the scores are reasonably represented as a Gaussian random variable.

This analysis was performed for the 46 stroke play tournaments held as part of the 2007 PGA Tour from January 4 thru November 4, (including the Masters Tournament, United States Open Championship, British Open Championship, and PGA Championship) using the scores reported on the PGA tour web site, www.pgatour.com. For each event, the first and second moments are calculated, yielding mean and variance. The probability distribution of the scores is calculated and the resulting *CDF* is compared in a *K-S* test to the *CDF* of the corresponding model function. This analysis results in a

*p*-value for each of the 46 tournaments. The *CDF* of this distribution of *p*-values is indicated in Fig. 3 as the crosses. While the great majority of *p*-values are $p > 0.7$, there is a significant tail that extends to values $p < 0.2$.

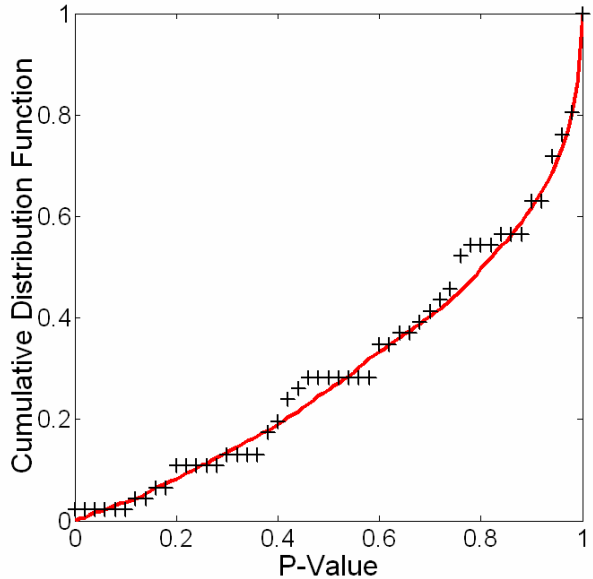

Fig. 3: Shown as crosses is the cumulative distribution of *p*-values obtained by performing a *K-S* analysis for each of the 46 tournaments, comparing the scores to the model function. The solid line is the cumulative distribution of *p*-values obtained by simulating the results of the 46 tournaments using the mean, standard deviation, and total number of scores for the actual events; assuming the scores are distributed as a Gaussian random variable; and then iterating the process 100 times.

To understand this distribution of *p*-values, the following numerical model was considered. All 46 tournaments were simulated assuming the scores to be distributed as a Gaussian random variable with mean, standard deviation, and total number of scores corresponding to the actual events. The resulting sampling of scores were compared in a *K-S* test against the model function, which consists of $10^5$ Gaussian distributed samples of the same mean and standard deviation. This process was then iterated 100 times, *i.e.* as if running the PGA tour for 100 years, resulting in 4600 *p*-values. The *CDF* of these

*p*-values is shown as the solid line in Fig. 3. One can use a *K-S* analysis to compare the *CDF* of this model distribution of *p*-values with the *CDF* of *p*-values obtained from analysis of the 46 tournaments. This analysis yields *p* ~ 0.80, making it very difficult to reject the null-hypothesis that <u>all scores on the PGA tour are drawn from Gaussian distributions</u>.

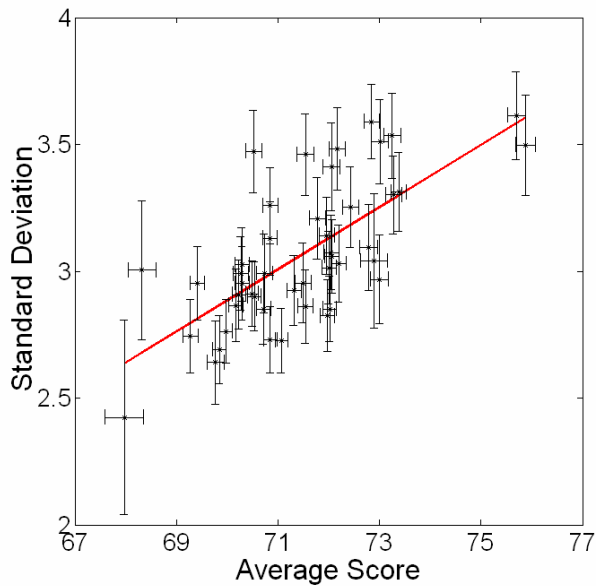

Fig. 4: The average score (abscissa) plotted parametrically against the standard deviation (ordinate) for each of the 46 stroke play tournaments. The error bars are estimates of the uncertainty, as described in the text. The solid line is a linear fit of the data, yielding a slope of 0.12. This indicates that more difficult golf courses do a slightly better job of separating the better players from the poorer players.

The strength of this analysis is that one can separate the performance of the field of competitors from the difficulty of the venue. As a first example, shown in Fig. 4 is the mean score ($\mu_s$) plotted vs. the standard deviation of the scores ($\sigma_s$) for all 46 tournaments. The uncertainties are represented as error bars and estimated as $\sigma_s/\sqrt{N}$, where *N* is the number of scores reported for each tournament. Note that the Masters Tournament and the U.S. Open Championship have anomalously high mean scores, 75.1 strokes and 76.2 strokes respectively. The solid line is a linear fit to the data and has a slope $\Delta\sigma_s/\Delta\mu_s = 0.12$. This line highlights the trend of increasing $\sigma_s$ with increasing $\mu_s$, indicating that harder courses do a little better at separating the better players from the rest of the field.

The most popular methods for comparing the overall performance of competitors on the PGA tour are the money list and the scoring average. This analysis suggests an alternative method for comparing performance, where in one keeps track of the *z*-scores for each competitor. The *z*-score, sometimes called either standard score or normal score, is a dimensionless measure of the difference from the mean in terms of the number of standard deviations. It is calculated by subtracting the mean from the raw score and dividing by the standard deviation. The *z*-score is a standard statistical tool used for comparing observations from different normal distributions [8]. The *z*-score methodology applied to scores on the PGA tour would allow one to understand how individual players perform relative to the field, independent of the difficulty of the courses. Additionally, the z-score characterizes performance in terms of a statistic that has quantitative meaning. Several examples of the uses of this methodology are provided in the following paragraphs.

Z-scores were calculated for all players returning scores in the 46 events on the 2007 PGA tour. The average *z*-score, $\mu_z$, for each player was then calculated, along with the corresponding standard deviation of the *z*-score, $\sigma_z$, for each player. These $\mu_z$ are plotted in Fig. 4 as a function of the player's position on the 2007 PGA Tour money list for the top 200 players on the money list. The star indicates $\mu_z$ and the vertical error bar indicates the uncertainty in our estimate of $\mu_z$, which we approximate as $\sigma_z/\sqrt{N}$, $N$ being the total number of scores reported for the player in 2007. There are several interesting things to note in this graphic. First, the resulting curve is linear with a slope $0.0023/\text{position}$. If one assumes $\sigma_s \approx 3$ strokes, (see Fig. 4), then one concludes only 0.34 strokes per round separate a 50 place difference in position on the money list. Second, the cut at the 125th player for continued exemption on the PGA tour occurs very near to $\mu_z = 0$. Thus, competitors playing better than the average of the field will likely keep their PGA tour exemption. Finally, $\mu_z$ for the first player on the money list, Tiger Woods, stands alone. His $\mu_z$ is approximately 1.05, while his closest competitors have $\mu_z \sim 0.5$. This amounts to Mr. Woods typically being in the top 15% of the field on any given day, while his closest competitors are typically in the top 30% of the field.

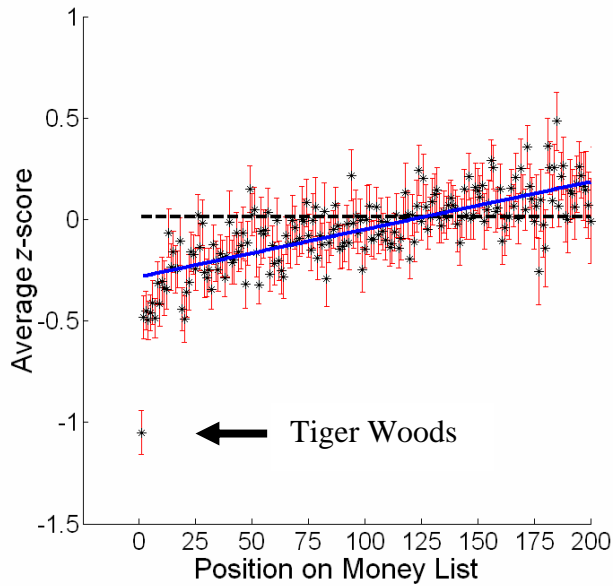

Fig 5: The vertical bars and stars represent the average $z$-score for each of the top 200 players on the 2007 PGA tour money list. The solid line is a linear fit to the z-scores of all but the first ranked player on the money list. The slope of the solid line is $0.0023/\text{position}$. The dashed line is the value of $\mu_z$ corresponding to the 125th position on the money list, which corresponds very nearly to $\mu_z = 0$. The number one ranked player on the money list, Tiger Woods, has $\mu_z \sim 1.05$, which is anomalously low in comparison to all other players.

Another use of the $z$-score is to track the performance of individual players over time. The $z$-scores for each player on the 2007 PGA tour were charted chronologically. A linear fit was then performed on the data as a means of identifying trends. Based on this analysis, Justin Leonard was the most improved player of the top 125 players on the money list. His $z$-scores are shown in Fig. 6, plotted chronologically. His average $z$-score $\mu_z$ is $-0.12$, and is indicated by the dashed line. The linear trend, indicated as the solid line, suggests he improved by almost a full standard deviation, from $\mu_z = 0.41$ to $\mu_z = -0.62$ over the course of the year. Assuming $\sigma_s \approx 3$, this amounts to an improvement of three strokes per round over the course of the year.

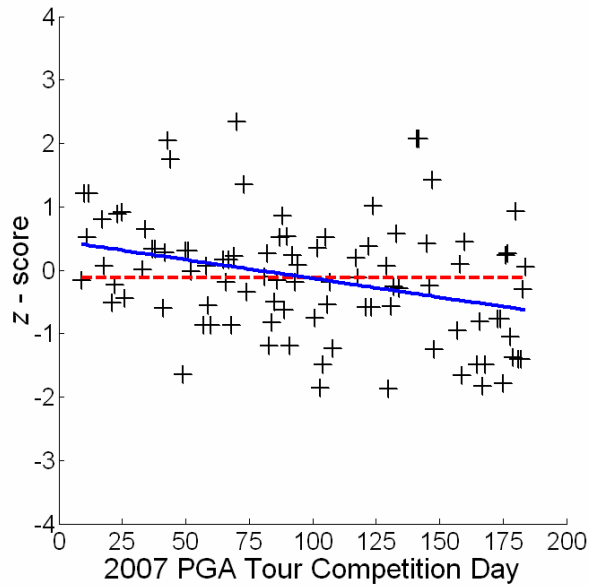

Fig 6: The $z$-scores for Justin Leonard, arranged chronologically, for all rounds he played on the 2007 PGA tour. His average $z$-score is $\mu_z = -0.12$, and is indicated by the dashed line. The linear trend, indicated as the solid line, suggests that he improved by almost a full standard deviation, from $\mu_z = 0.41$ to $\mu_z = -0.62$ over the course of the year. Based on this analysis, Justin Leonard was the most improved player of the top 125 players on the 2007 money list.

It is worth noting that the third most improved player in 2007 was Tiger Woods. His z-scores are charted chronologically in Fig. 7. While his average z-score $\mu_z$ is $-1.05$, the linear fit suggests he improved by 2/3 of a standard deviation, from $\mu_z = -0.64$ to $\mu_z = -1.33$, over the course of the year. This amounts to an improvement of two shots per round. Relevant to the following analysis, it is notable that he often has clusters of z-scores less than or equal to $-1.5$, and occasionally has clusters exceeding $-2.0$.

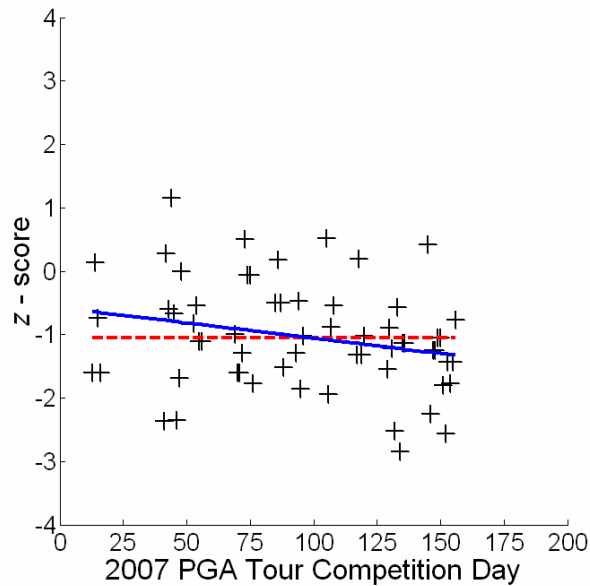

Fig 7: Z-scores for all 2007 scores reported for Tiger Woods, arranged chronologically. While his average z-score $\mu_z$ is $-1.05$, the linear fit suggests that he improved by 2/3 of a standard deviation, from $\mu_z = -0.64$ to $\mu_z = -1.33$, over the course of the year. Based on this analysis, Tiger Woods was the third most improved player of the top 125 players on the 2007 money list.

Finally, one can use the z-scores in a statistical analysis to predict the outcome of tournaments. As an example, we have calculated the probability that a fictitious player will tie or break Byron Nelson's long standing record of eleven consecutive PGA Tour victories over the course of a 300 tournament career. This calculation was chosen so as to inform the ongoing speculation as to whether Tiger Woods will break this record in the remaining years of his career. For the purposes of this calculation, his remaining career is assumed to span 300 tournaments, *i.e.* 20 tournaments per year for the next 15 years.

The calculation is as follows. The field of competitors is defined by the ensemble of $\mu_z$ and $\sigma_z$ indicated in Fig. 5, excluding those of Tiger Woods. A tournament is simulated by assuming the players comprise the top 155 competitors in the field plus a fictitious competitor. This fictitious competitor is assigned a value of $\mu_z$ and $\sigma_z$. The particular value of $\sigma_z$ is that of Tiger Woods, which is calculated using the data in Fig. 7. As is discussed below, $\mu_z$ for this fictitious player is fixed for the duration of the career, and various different careers are modeled by changing the value of $\mu_z$. Using this ensemble of $\mu_z$ and $\sigma_z$, and a Gaussian random number generator [9], a tournament is simulated by generating four scores for each player. The total for each player is calculated, the result of the virtual tournament is tabulated, and it is determined if the fictitious competitor wins the event.

The career of this fictitious competitor is modeled as the results for 300 simulated tournaments. The total number of wins and the maximum number of consecutive wins for the competitor is tabulated. Each career is then run $10^4$ times to improve the resulting statistics. This is then done for various values of $\mu_z$, so as to gauge how the results vary as a function of $\mu_z$.

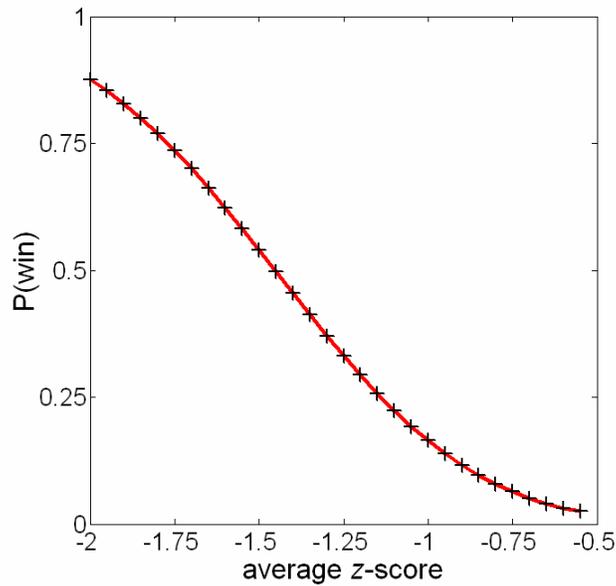

Figure 8: The probability of our fictitious competitor winning a tournament as a function of $z$-score. Details of the calculation are described in the text. The crosses indicate the calculated data points. The solid line is drawn as an aid to the eye. Note that the probability approaches 0.5 as $\mu_z$ approaches $-1.5$.

The graphic in Fig. 8 indicates the probability of victory in a tournament for the fictitious competitor as a function of $\mu_z$. The data are indicated by the crosses. The solid line connects the data points as an aid to the eye. The probability of victory is approximately 1 in 40 for $\mu_z = -0.5$, rising to approximately 1 in 2 for $\mu_z = -1.5$.

The graphic in Fig. 9 indicates the probability that the fictitious competitor wins eleven or more consecutive tournaments over the course of a career, as a function of $\mu_z$. The data are indicated by the crosses. The solid line connects the data points as an aid to the eye. The graphic shows the probability to be negligible for $\mu_z$ as low as $-1.5$. In order to have even odds of achieving eleven or more consecutive victories, the fictitious competitor requires a value of $\mu_z$ approaching $-2.0$ over the course of the entire 300 tournament career. As shown in Fig. 7, Mr. Woods occasionally has periods of play with

$\mu_z$ as low as -2.0; however, the existing data does not support the notion that this level of performance can be maintained over the course of a 15 year career. At the very least, it provides Mr. Woods with a challenging and quantitative goal.

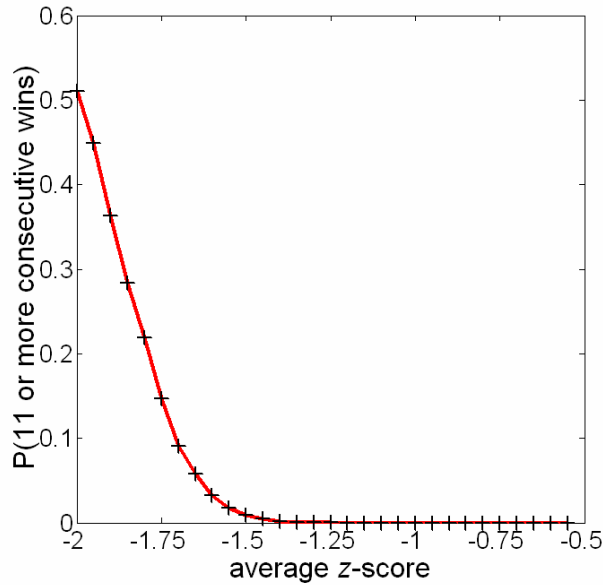

Figure 9: The probability of eleven or more consecutive wins as a function of $\mu_z$ over the course of a 300 tournament career. Details of the calculation are described in the text. The crosses indicate the calculated data points. The solid line is drawn as an aid to the eye. Note that the probability approaches even odds as the value of $\mu_z$ approaches $-2.0$.

In summary, it has been demonstrated that the scores generated on the PGA tour are reasonably modeled in terms of Gaussian statistics. This analysis suggests the *z*-score is a valuable alternative methodology for measuring performance on the PGA tour, as it provides a means of measuring performance relative to the field of competitors that is independent of the relative difficulty of the various tournament golf courses. As an example of this methodology, the *z*-score was used to identify the most improved player on the 2007 PGA tour to be Justin Leonard. Additionally, the *z*-scores are used to simulate the career of a fictitious player in an attempt to inform speculation as to the likelihood Tiger Woods will be able to equal or better Byron Nelson's record of eleven consecutive PGA tour victories. This analysis suggests a player must maintain an average *z*-score of order $-2.0$ for the duration of a 15 year career in order to have even odds of breaking this record; a feat that seems extraordinarily challenging.

Acknowledgements:

The author acknowledges useful conversations with and thoughtful advice from Professor Joseph Chang of Yale University and Professor William Press of University of Texas at Austin.